\documentclass[preprint,proceedings]{rmaa}

\suppressfulladdresses % by popular request

\usepackage{paralist}

\usepackage{psfrag,color}

\SetYear{2007}
\SetConfTitle{Massive Stars: Fundamental Parameters and Circumstellar Interactions 
(Workshop to celebrate the 70th birthday of Professor Virpi Niemela)}

\title{Magnetic fields in massive stars} 

\author{
  S. Hubrig,\altaffilmark{1}} 
  
\altaffiltext{1}{European Southern Observatory, Casilla 19001, Santiago 19, Chile (shubrig@eso.org).}

\shortauthor{Hubrig}
\shorttitle{Magnetic fields in massive stars}

\listofauthors{S. Hubrig}
\indexauthor{Hubrig, S.}
\abstract{
Although indirect evidence for the presence of magnetic fields in high-mass stars is 
regularly reported in the literature, the detection of these fields remains an extremely 
challenging observational problem. We review the recent discoveries of magnetic fields in 
different types of massive stars and briefly discuss strategies for 
spectropolarimetric observations to be carried out in the future.
}

\addkeyword{Stars: Circumstellar matter}
\addkeyword{Stars: Magnetic fields}
\addkeyword{Stars: Oscillation}
\addkeyword{Stars: Polarisation}
\begin{document}
% Typeset article header
\maketitle

\section{Introduction}
\label{sec:intro}

An increasing number of observations of hot stars provide indirect evidence that 
magnetic fields must be present in those stars: Cyclic behaviour on a rotational 
timescale observed in the UV wind lines (e.g., Henrichs et al.\ 2005), the presence of narrow X-ray 
emission lines (Cohen et al.\ 2003), non-thermal radio emission (Bieging et al.\ 1989), etc.
In spite of numerous indirect evidence only very few direct magnetic field detections 
have been reported so far. Magnetic fields are accessible through the Zeeman effect:
The Zeeman components of spectral lines are polarised and thus permit magnetic fields to be 
measured even in rapidly rotating massive stars where rotation broadening, etc., prevents 
the resolution of Zeeman components.

\section{Recent searches}
\label{sec:rece}

Currently, direct measurements are achieved only in two O stars, $\theta^1$\,Ori\,C and HD\,191612 
with B$_{\rm eff}$ values of a few hundred Gauss (Donati et al.\ 2002, 2006). To study the 
incidence of magnetic fields in O stars, we recently obtained high S/N spectropolarimetric 
observations of 
eleven O stars with FORS\,1 at the VLT with a typical accuracy of the field determination of 
about 30--70\,G. However, no evidence of a magnetic field was found, leading to the conclusion 
that large scale, dipole like magnetic fields are not widespread among O-type stars.

\begin{figure}[!t]
  \includegraphics[width=\columnwidth]{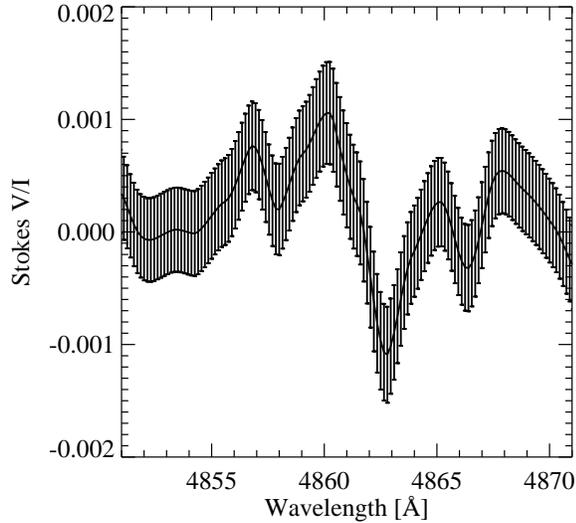}
  \caption{Stokes V/I spectra of the $\beta$\,Cephei star $\xi^1$\,CMa around 
the H$\beta$ line. The thickness of the plotted line corresponds to the uncertainty 
of the polarisation measurements determined from photon noise.}
  \label{hubrig:hbeta}
\end{figure}

\begin{figure}[!t]
  \includegraphics[width=\columnwidth]{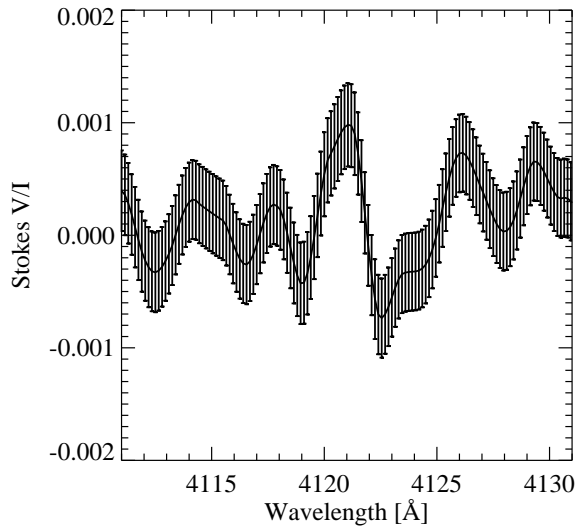}
   \caption{Stokes V/I spectra of the $\beta$\,Cephei star $\xi^1$\,CMa around 
the He~II line at 4121\,\AA{}.}
  \label{hubrig:he}
\end{figure}

Among early B-type stars, a magnetic field has been discovered in the B0.2V star $\tau$\,Sco and 
in one of the hottest $\beta$\,Cephei stars, the B0.7IV star $\xi^1$\,CMa. Two more 
$\beta$\,Cephei stars, $\beta$\,Cep (Henrichs et al.\ 2000) and V2053\,Oph (Neiner et al.\ 
2003a), with spectral types B2III and B2IV, 
exhibit a weak magnetic field. The star $\xi^1$\,CMa has the largest magnetic field
of up to 300\,G (Hubrig et al.\ 2006), whereas the magnetic field for the other three stars is much 
weaker with  corresponding B$_{\rm eff}$ values less than 100\,G. 
A magnetic field of the order of a few hundred Gauss has recently been discovered 
for slowly pulsating B-type stars (Neiner et al.\ 2003b, Hubrig et al.\ 2006).
The effect of the magnetic field on the oscillation 
properties of $\beta$\,Cephei stars and slowly pulsating B-type stars has not been studied in detail yet.
In Figs.~\ref{hubrig:hbeta} and \ref{hubrig:he} we present the most recent magnetic field observations of the 
hottest $\beta$\,Cephei star $\xi^1$\,CMa.

Another type of massive stars, rapidly rotating Be stars, lose mass and initially 
accumulate it in a rotating circumstellar disk. Much of the mass loss is in the form 
of outbursts and so additional mechanisms such as the beating of nonradial 
pulsation modes or magnetic flares must be at work. Previously, only two Be stars 
have been found as weekly magnetic with B$_{\rm eff}$ values less than 
100\,G (Neiner et al.\ 2003c, Hubrig et al.\ 2006).
Our recent search for magnetic fields in 15 Be stars with FORS\,1 at the VLT revealed 
that out of our sample five Be stars present circular polarisation signatures in lines formed 
in the circumstellar environment (Hubrig et al.\ 2007). Furthermore, for a few stars we measured a field 
of the order of 100\,G, changing its field polarity every 8--9 minutes. Clearly, time series 
are needed to study these local transient magnetic fields. As it has already been discussed 
by Maheswaran (2003) in the framework of his Magnetic Rotator Wind-Disk Model,
magnetic fields can consist 
of flux loops that emerge from lower latitudes and thread the disk around the Be star.

\section{Discussion}
\label{sec:disc}

Magnetic fields are indeed present in hot massive stars. Potential progress in their study 
may come from achieving better accuracy of the measurements of magnetic fields or from the detailed 
studies of polarised line profiles, in the case that the magnetic fields of massive stars 
have structures significantly more complex than those of classical Ap and Bp stars.

The Hanle effect is a relatively new magnetic diagnostic in stellar astrophysics, but 
can probably be used to study circumstellar magnetic fields.
This effect produces a depolarisation of the scattered radiation and a rotation of 
the plane of polarisation (e.g., Ignace et al.\ 2004). A program for such observations is 
currently allocated at the 
VLT with FORS\,1 and we are expecting to receive the data in the middle of 2007.


\begin{thebibliography}

\bibitem{bi89} Bieging, J.~H., Abbott, D.~C., \& Churchwell, E.~B.\ 1989,
ApJ, 340, 518

\bibitem{co03} Cohen, D.~H., et al.\ 2003, ApJ, 586, 495 

\bibitem{do02} Donati, J.-F., Babel, J., Harries, T.~J., Howarth, I.~D., Petit, P., \& Semel, M.\ 2002,
MNRAS, 333, 55

\bibitem{do06} Donati, J.-F., Howarth, I.~D., Bouret, J.-C., Petit, P., Catala, C., \& Landstreet, J.\ 2006,
MNRAS, 365, 6

\bibitem{he00} Henrichs, H.~F., et al.\ 2000,
in ASP Conference Series, Vol.\ 214, ``The Be Phenomenon in Early-Type Stars'',
eds.\ Smith, M.~A., \& Henrichs, H.~F., p.~324

\bibitem{he05} Henrichs, H.~F., Schnerr, R.~S., \& Ten Kulve, E.\ 2005,
in ASP Conference Series, Vol.\ 337, ``The Nature and Evolution of Disks Around Hot Stars'',
eds.\ Ignace, R.\ \& Gayley, K.~G., p.~114

\bibitem{hu06} Hubrig, S., Briquet, M., Sch\"oller, M., De Cat, P., Mathys, G., \& Aerts, C.\ 2006,
MNRAS, 369, 61

\bibitem{hu07} Hubrig, S., et al.\ 2007,
{\sl in preparation}

\bibitem{ign04} Ignace, R., Nordsieck, K.~H., \& Cassinelli, J.~P.\ 2004,
ApJ, 609, 1018

\bibitem{ma03} Maheswaran, M.\ 2003,
ApJ, 592, 1156

\bibitem{ne03a} Neiner, C., et al.\ 2003a,
A\&A, 411, 565

\bibitem{ne03b} Neiner, C., et al.\ 2003b,
A\&A, 406, 1019

\bibitem{ne03c} Neiner, C., et al.\ 2003c,
A\&A, 409, 275

\end{thebibliography}
\end{document}